\documentclass[twocolumn]{aastex62}

\newcommand{\feh}{$\mbox{[Fe/H]}$}

\received{--}
\revised{--}
\accepted{--}

\submitjournal{ApJ}

\shorttitle{Metal-poor stars in the Sagittarius dwarf galaxy}
\shortauthors{Chiti, Hansen, \& Frebel}

\begin{document}

\title{Discovery of 18 stars with $-3.10<$ [Fe/H] $< -1.45$ in the Sagittarius dwarf galaxy\footnote{This paper includes data gathered with the 6.5\,m Magellan Telescopes located at Las Campanas Observatory, Chile.}}

\correspondingauthor{Anirudh Chiti}
\email{achiti@mit.edu}

\author[0000-0002-7155-679X]{Anirudh Chiti}
\affil{Department of Physics and Kavli Institute for Astrophysics and Space Research, Massachusetts Institute of Technology, Cambridge, MA 02139, USA}

\author{Kylie Y. Hansen}
\affil{Department of Physics and Kavli Institute for Astrophysics and Space Research, Massachusetts Institute of Technology, Cambridge, MA 02139, USA}

\author[0000-0002-2139-7145]{Anna Frebel}
\affiliation{Department of Physics and Kavli Institute for Astrophysics and Space Research, Massachusetts Institute of Technology, Cambridge, MA 02139, USA}

\begin{abstract}

Studies of the early chemical evolution of some larger dwarf galaxies ($> 10^7$\,M\textsubscript{\(\odot\)}) are limited by the small number of stars known at low metallicities in these systems.
Here we present metallicities and carbon abundances for eighteen stars with metallicities between $-3.08 \le \text{[Fe/H]} \le -1.47$ in the Sagittarius dwarf spheroidal galaxy, using medium-resolution spectra from the MagE spectrograph on the Magellan-Baade Telescope. 
This sample more than doubles the number of known very metal-poor stars ([Fe/H] $\leq -2.0$) in the Sagittarius dwarf galaxy, and identifies one of the first known extremely metal-poor stars ([Fe/H] $\leq -3.0$) in the system.  
These stars were identified as likely metal-poor members of Sagittarius using public, metallicity-sensitive photometry from SkyMapper DR1.1 and proper motion data from \textit{Gaia} DR2, demonstrating that this dearth of metal-poor stars in some dwarf galaxies can be addressed with targeted searches using public data.
We find that none of the stars in our sample are enhanced in carbon, in contrast to the relative prevalence of such stars in the Milky Way halo.
Subsequent high-resolution spectroscopy of these stars would be key in detailing the early chemical evolution of the system.

\end{abstract}


\section{Introduction} 
\label{sec:intro}

The Milky Way's metal-poor stars\footnote{Defined as $\feh\,\,\le -1\,\text{dex}$, where [Fe/H] = $\log_{10}(N_{\text{Fe}}/N_{\text{H}})_{\star}-\log_{10}(N_{\text{Fe}}/N_{\text{H}})_{\sun}$ \citep{bc+05, fn+15}.} are a nearby window to the high-redshift universe. 
The natal chemical composition of these ancient stars is generally preserved in their stellar atmospheres.
Accordingly, their chemical abundances trace the formation of the first heavy elements (see review \citealt{fn+15}) and can be used to infer the properties (e.g., initial mass function, explosion energy, chemical yields) of the First Stars and supernovae that drove early nucleosynthesis \citep{hw+10, itk+18}.
The chemical abundances of metal-poor stars also reflect the early evolution of their host galaxies, since an interplay of several galactic properties, including star formation efficiency, accretion, and feedback, drives chemical evolution \citep{ths+09, rs+13, kcg+13}.

In this context, the Milky Way's satellite dwarf galaxies are prime environments to identify and chemically characterize metal-poor stars to learn about early galactic environments. 
These systems span a range of masses ($\sim10^5$\,\(\textup{M}_\odot\) to $\sim10^{11}$\,\(\textup{M}_\odot\); e.g., \citealt{wmo+16,s+19, ebl+19}) and thus allow the investigation of early chemical evolution in a variety of self-contained environments.
Furthermore, analogs of these surviving satellite dwarf galaxies may have been accreted onto the Milky Way to form the old stellar halo.
Accordingly, diversity in the chemical abundances of very metal-poor ([Fe/H $< -2.0$) stars in the Milky Way halo may be explained by its assembly from smaller galaxies \citep{dmw+16, bjf+19}.
To fully test this claim, and then consequently isolate the environments that produce the chemical signatures we observe in the oldest, most metal-poor stars, it is necessary to identify large samples of similarly metal-poor stars in the Milky Way's satellite dwarf galaxies. 

However, the most metal-poor population ([Fe/H] $\lesssim -2.5$) of the most massive Milky Way dwarf spheroidal (dSph) galaxies ($>10^7$\,\(\textup{M}_\odot\)) remains sparsely characterized, as galaxies with larger masses have much higher average metallicities \citep{kcg+13}.
Only 55 stars with [Fe/H] $< -2.5$ known across all dSphs have detailed chemical abundance measurements available \citep{scs+01, frc+04, aas+09, aaf+20, ch+09, ch+10, fks+10, tjh+10, kc+12,  vsi+12, uck+15, sjf+15, jnm+15, hem+18, tjl+19}, with the majority in just two systems: the Sculptor and Sextans dSphs. 
Identifying more such metal-poor stars across all dSphs is informative when addressing the aforementioned questions, especially since these larger dwarf galaxies are thought to have contributed the most stars to the Milky Way halo \citep{dmw+16}.

Here we present the results of a targeted search for very metal-poor stars in the Sagittarius dSph \citep{igi+94,igi+95}.
Sagittarius is the most massive known satellite dwarf spheroidal galaxy of the Milky Way \citep[$\sim4\times10^8$\,\(\textup{M}_\odot\);][]{vv+20}. It has only a handful of known very metal-poor stars \citep{bic+08, mbi+17, hem+18, cf+19}, due to a prominent metal-rich ($\langle$[Fe/H]$\rangle$ $\sim -0.5$) component of its stellar population \citep{mbi+17}.
A detailed abundance analysis of just three of these very metal-poor stars in \citet{hem+18} already hinted at some similarities to the chemical abundances of very metal-poor stars in the Milky Way halo. It also challenged claims of a top-light initial mass function as suggested by its more metal-rich population \citep[e.g.,][]{mwm+13}.
To further investigate the early evolution of this system, we present metallicities and carbon abundances for eighteen red giant metal-poor stars in the Sagittarius dSph that were identified using metallicity-sensitive SkyMapper photometry \citep{ksb+07,wol+18} and Gaia DR2 proper motions \citep{gaia+16,gaia+18} following \citet{cf+19}.
Notably, we identify nine stars with [Fe/H] $\leq -2.0$, one of which is one of the first known extremely metal-poor stars ([Fe/H] $< -3.0$) in the system\footnote{While under consideration, we became aware of a recent homogenous re-analysis of dSph stars in \citet{rhh+20} that had also recovered an extremely metal-poor star in Sgr\label{ft:sgr}.}. 

This paper is organized as follows. 
In Section~\ref{sec:obs}, we provide an overview of our observations; 
in Section~\ref{sec:analysis}, we outline our methodology in deriving stellar parameters and chemical abundances.
We discuss the implications of our results on the early evolution of the Sagittarius dSph in  Section~\ref{sec:results}, and summarize our work in Section~\ref{sec:results}.

%%%%%%%%%%%%%%%%%%%%%%%%%%%%%%%%%%%%%%%%%%%%%%%%%%%%%%%%%%%%%%%%%%%%%%%%%%%%%%%%%
\begin{deluxetable*}{lllrrrrr}[!htbp] % <--- column justification (center/left/right)
\tablecaption{\label{tab:obs} Observations}
\tablehead{   % column headings
  \colhead{Name} &
  \colhead{RA (h:m:s)} & 
  \colhead{DEC (d:m:s)} &
%  \colhead{Slit size} &
  \colhead{$g$ } &
  \colhead{$t_{\text{exp}}$ } &
  \colhead{S/N$\tablenotemark{a}$} & 
  \colhead{$v_{\text{helio}}$ } \\
   \colhead{}&
   \colhead{(J2000)}&
   \colhead{(J2000)}&
%   \colhead{}&
   \colhead{[mag]}&
   \colhead{[s]}&
   \colhead{}&
   \colhead{[km\,s$^{-1}$]}&
  }
\startdata
Sgr-300 & 18:44:26.84 & $-$29:37:56.09 & 15.82 & 480 & 40,85 & 132 \\
Sgr-265 & 18:44:56.86 & $-$31:12:01.84 & 16.09 & 480 & 40,75 & 165 \\
Sgr-180 & 18:48:43.24 & $-$31:46:26.82 & 15.66 & 420 & 25,55 & 152 \\
Sgr-157 & 18:48:51.47 & $-$31:35:21.58 & 17.30 & 1200 & 30,40 & 133 \\
Sgr-298 & 18:49:32.88 & $-$32:44:26.81 & 17.46 & 1200 & 30,40 & 165 \\
Sgr-91 & 18:51:44.32 & $-$29:30:38.85 & 16.17 & 540 & 40,65 & 125 \\
Sgr-69 & 18:52:48.45 & $-$29:32:23.42 & 15.56 & 480 & 45,95 & 129 \\
Sgr-48 & 18:56:26.25 & $-$31:21:23.61 & 15.83 & 480 & 45,85 & 142 \\
Sgr-81 & 18:56:52.64 & $-$31:43:07.51 & 17.39 & 900 & 25,40 & 138 \\
Sgr-38 & 18:57:27.66 & $-$31:07:39.87 & 16.26 & 540 & 40,65 & 150 \\
Sgr-139 & 18:57:50.62 & $-$29:00:29.93 & 17.35 & 1200 & 40,50 & 119 \\
Sgr-198 & 18:57:51.93 & $-$28:37:08.95 & 15.95 & 480 & 25,70 & 138 \\
Sgr-141 & 18:59:07.78 & $-$29:08:15.52 & 16.31 & 540 & 35,60 & 118 \\
Sgr-62 & 18:59:13.41 & $-$31:12:39.45 & 16.16 & 480 & 35,60 & 157 \\
Sgr-182 & 19:00:50.31 & $-$29:04:53.46 & 17.41 & 1200 & 35,50 & 117 \\
Sgr-136 & 19:00:53.93 & $-$29:28:38.09 & 15.68 & 480 & 45,90 &  155 \\
Sgr-162 & 19:02:12.05 & $-$31:29:37.94 & 17.40 & 1200 & 40,60 & 156 \\
Sgr-215 & 19:04:30.07 & $-$29:56:14.05 & 16.26 & 540 & 45,55 & 130 \\
Sgr-225 & 19:05:06.80 & $-$30:19:22.89 & 16.00 & 480 & 45,75 & 120 \\
Sgr-333 & 19:07:29.65 & $-$29:58:01.35 & 16.08 & 480 & 40,70 & 126 \\
\enddata
\tablenotetext{a}{S/N per pixel is listed both for 4500\,\AA\,\,and 8500\,\AA}
\end{deluxetable*}
%%%%%%%%%%%%%%%%%%%%%%%%%%%%%%%%%%%%%%%%%%%%%%%%%%%%%%%%%%%%%%%%%%%%%%%%%%%%%%%%%

\section{Target Selection \& Observations}
\label{sec:obs}

We identified metal-poor candidate members of the Sagittarius dwarf spheroidal galaxy (dSph) following the criteria in \citet{cf+19}, which we briefly outline here.
We retrieved photometric data from the SkyMapper DR1.1 catalog \citep{wol+18} on all sources within three degrees of the center of the Sagittarius dSph.
This dataset includes photometric data obtained through the metallicity-sensitive SkyMapper $v$ filter \citep{bbs+11,dbm+19}, which is also sensitive to surface gravity when compared to photometry from the SkyMapper $u$ filter \citep{mks+09}.
We thereby derived metallicities and surface gravities for all of these sources using the methods described in \citet{cfj+20} and identified likely metal-poor giants ([Fe/H] $< -2.0$ and $\log\,g < 3.0$) for observations.
We note that the photometric metallicities for these stars had large uncertainties ($\sim0.75$\,dex) due to the large uncertainties on the SkyMapper $u$ and $v$ photometry at the magnitudes ($g\sim15.5$ to $g\sim17.5$) of these stars.
As a result, we regarded these metallicity and surface gravity cuts  effectively qualitative selection criteria.

\begin{figure}[!htbp]
\centering
\includegraphics[width =\columnwidth]{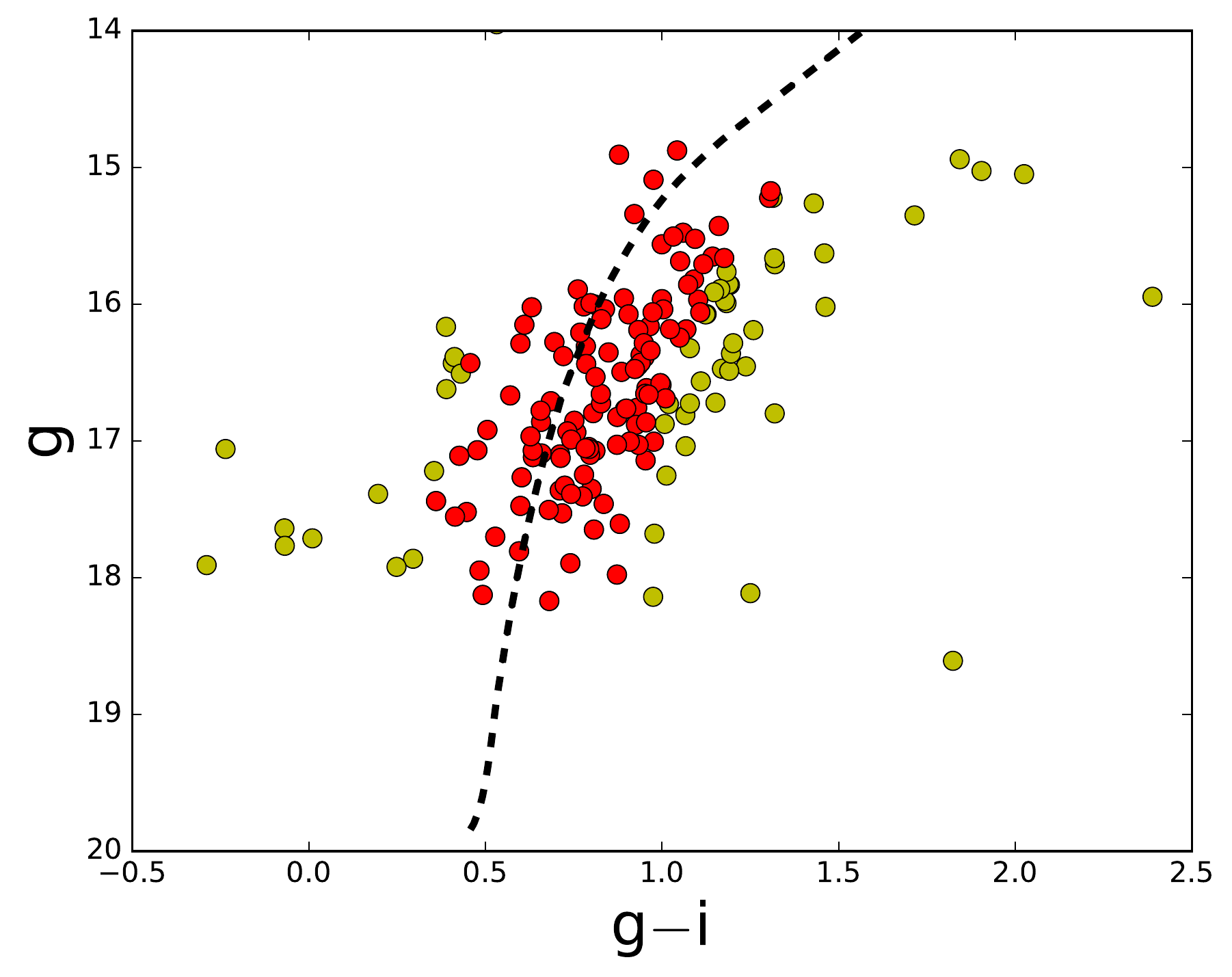}
\caption{A color-magnitude diagram of all stars within three degrees of the center of the Sagittarius dSph that pass the photometric metallicity, surface gravity, and proper motions criteria listed in Section~\ref{sec:obs}.
A 10\,Gyr, [Fe/H] = $-2.0$ isochrone from the Dartmouth Stellar Evolution Database \citep{dcj+08} is over plotted for reference. 
Stars in red are within $(g-i)\pm0.30$ of the isochrone, and stars in yellow are outside those bounds.}
\label{fig:targ_iso}
\end{figure}

To increase the likelihood of membership, we selected the subset of these metal-poor giants with \textit{Gaia} DR2 proper motions \citep{gaia+16,gaia+18} near the systemic proper motion of the Sagittarius dSph. 
The proper motion ranges of $-4.2\,\text{mas/yr} < \mu_\alpha < -1.90\,\text{mas/yr}$ and $-2.15\,\text{mas/yr} < \mu_\delta < -0.85\,\text{mas/yr}$ were used to select likely members for the first night of observations. 
These criteria were narrowed to $-3.6\,\text{mas/yr} < \mu_\alpha < -2.70\,\text{mas/yr}$ and $-1.6\,\text{mas/yr} < \mu_\delta < -1.2\,\text{mas/yr}$ on subsequent nights to ensure a purer sample of members, since a number of non-members were observed near the bounds of our original proper motion cut.
We then generated a color-magnitude diagram of these remaining candidates using SkyMapper $g$ and $i$ photometry. 
We chose to observe candidates that were within $(g-i)\pm0.30$ of a 10\,Gyr, [Fe/H] = $-2.0$ isochrone from the Dartmouth Stellar Evolution Database \citep{dcj+08} placed at the distance modulus of the Sagittarius dSph (16.97; \citealt{kc+09}).
A color-magnitude diagram of stars fulfilling these selection criteria is shown in Figure~\ref{fig:targ_iso}.

%%%%%%%%%%%%%%%%%%%%%%%%%%%%%%%%%%%%%%%%%%%%%%%%%%%%%%%%%%%%%%%%%%%%%%%%%%%%%%%%%
%%%%%%%%%%%%%%%%%%%%%%%%%%%%%%%%%%%%%%%%%%%%%%%%%%%%%%%%%%%%%%%%%%%%%%%%%%%%%%%%%

We obtained spectra of 37 of these metal-poor candidate members using the Magellan Echellette (MagE) Spectrograph \citep{mbt+08} on the Magellan-Baade telescope during the nights of August 3-5, 2019.
Targets were observed with the 0\farcs7 slit and 1x1 binning, granting a resolution of $R\sim6700$ over a broad wavelength range of 3200\,\AA\,\, to 10000\,\AA.
The seeing was excellent ($\sim$0\farcs6) throughout these observations.
A spectrum of a ThAr calibration arc lamp was obtained after slewing to each target for wavelength calibration purposes. 
Our observations were reduced using the Carnegie Python pipeline \citep{k+03}\footnote{https://code.obs.carnegiescience.edu/mage-pipeline}.

Twenty-one stars that we observed were determined to be members of the Sagittarius dSph (see Section~\ref{sec:rvs}).
Of these members, we excluded the following from further analysis: one spectroscopic binary system and two stars with distorted H$\alpha$ absorption lines. 
This resulted in 18 members for which we derived chemical abundances. 
Details of the observations of these 18 members are provided in Table~\ref{tab:obs}.

\section{Analysis}
\label{sec:analysis}

\subsection{Radial Velocity Measurements}
\label{sec:rvs}

We derived radial velocities by cross-correlating our spectra with a template spectrum of the metal-poor giant HD122563 that was obtained on the second night of data collection.
The rest velocity of HD122563 was assumed to be $-$26.51\,km\,s$^{-1}$ \citep{cmf+12}, and the cross-correlation was performed over the Mg b absorption region (4900\,\AA\,\,to 5400\,\AA).
We further performed a cross-correlation over the Telluric A-band absorption region (7590\,\AA\,\,to 7710\,\AA) between our spectra and a template spectrum of the hot, rapidly rotating giant HR4781 to correct for any slit mis-centering.
We note that this template spectrum of HR4781 was obtained with the IMACS instrument \citep{sld+17}, but was then smoothed to the resolution of our MagE spectra ($R\sim6700$).
We calculate heliocentric velocity corrections using the SkyCoord function in the Astropy package \citep{astropy}.
These methods resulted in a systematic velocity uncertainty of $\sim$3\,km\,s$^{-1}$, as determined by repeat observations of other dwarf galaxy stars in this observing mode (Chiti et al. 2020, subm.).
Table~\ref{tab:obs} lists our final velocity measurements.
 
\begin{figure}[!htbp]
\centering
\includegraphics[width =\columnwidth]{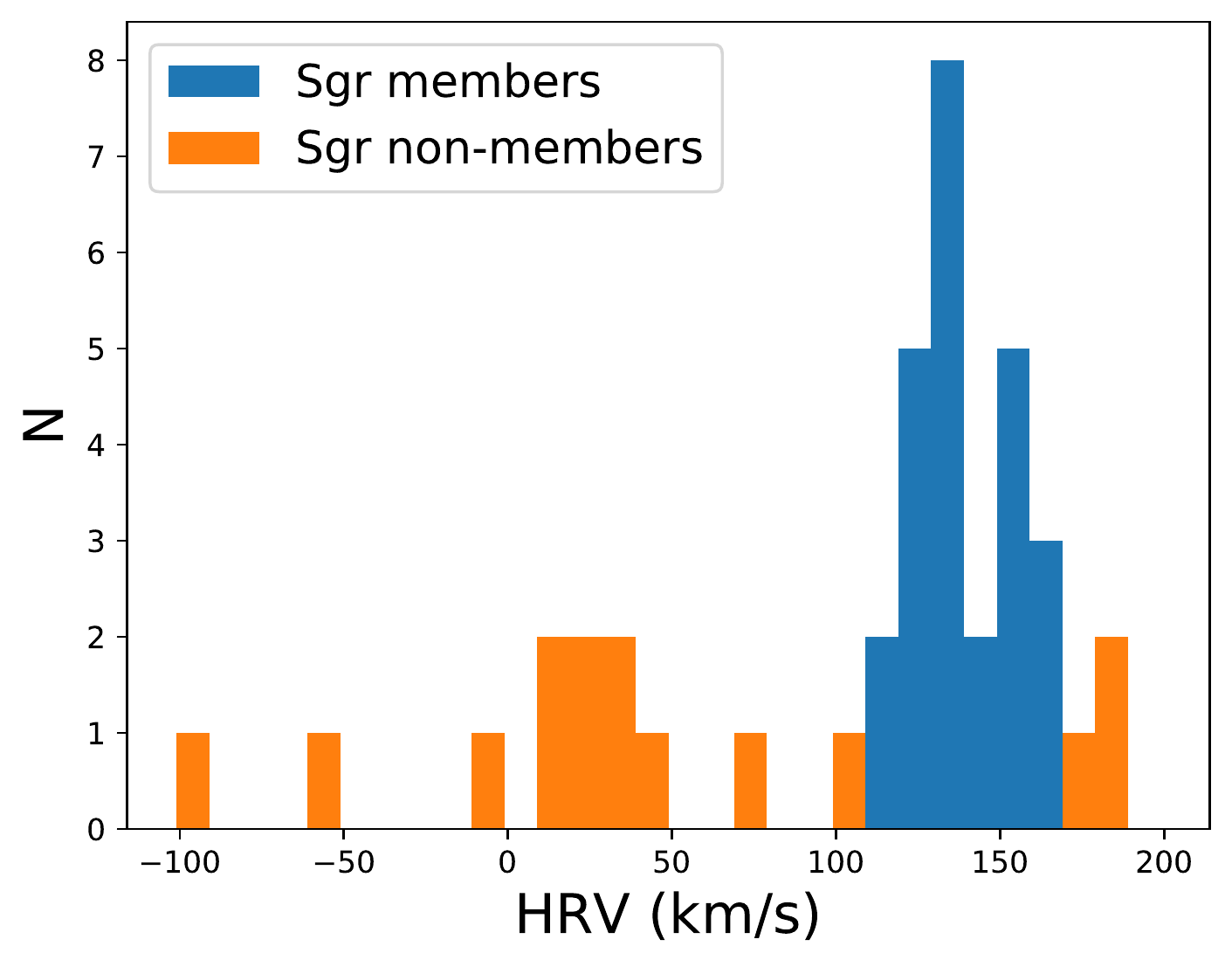}
\caption{Histogram of the heliocentric radial velocities of stars in our sample. 
The blue portion of the histogram represents the stars that we classify as members of the Sagittarius dSph, as determined by restricting radial velocity values between 111\,km\,s$^{-1}$ and 171\,km\,s$^{-1}$.}
\label{fig:vel_hist}
\end{figure}

We used these radial velocity measurements to determine whether the stars in our sample were members of the Sagittarius dSph.
The known stars in the bulge of the Sagittarius dSph display a systemic velocity of 141\,km\,s$^{-1}$ and a velocity dispersion of 9.6\,km\,s$^{-1}$ \citep{bic+08}. 
We find that the majority of our targets indeed lie within three times the velocity dispersion of this systemic velocity (111\,km\,s$^{-1}$ to 171\,km\,s$^{-1}$).
We identify those stars as members and classify the rest as nonmembers (see Figure~\ref{fig:vel_hist}).
The uncertainties in our velocity measurements are roughly equal to the systematic uncertainty ($\sim3$\,km\,s$^{-1}$), which is small relative to the velocity dispersion assumed for the Sagittarius dSph.

%%%%%%%%%%%%%%%%%%%%%%%%%%%%%%%%%%%%%%%%%%%%%%%%%%%%%%%%%%%%%%%%%%%%%%%%%%%%%%%%%
%%%%%%%%%%%%%%%%%%%%%%%%%%%%%%%%%%%%%%%%%%%%%%%%%%%%%%%%%%%%%%%%%%%%%%%%%%%%%%%%%

\subsection{Stellar Parameters}
\label{sec:stellpar}

We closely follow the methodology presented in \citet{cf+19} to derive the stellar parameters ($T_{\text{eff}},$ $\log{g}$) of the stars in our sample. 
Namely, we matched the SkyMapper $g-i$ colors of our stars to those colors in a 10\,Gyr, [Fe/H] = $-2.0$ isochrone from the MESA Isochrones \& Stellar Tracks database \citep{d+16, cdc+16, pbd+11, pca+13, pms+15, psb+18} and retrieved the corresponding stellar parameters. 
These colors were de-reddened following \citet{wol+18} using maps from \citet{sfd+98}.
As reported in \citet{cf+19}, applying this method on the sample of metal-poor Sagittarius dSph stars with already known stellar parameters derived from high-resolution spectroscopy in \citet{hem+18} results in stellar parameters that are in good agreement with those spectroscopic stellar parameters and thus validates this approach.
With our particular choice of isochrone, this method, on average, leads to a marginally larger $T_{\text{eff}}$ of 47\,K and $\log{g}$ of 0.4\,dex than those reported in \citet{hem+18}.
We note that these stellar parameter estimates are relatively insensitive to the assumed metallicity of the isochrone.
Increasing metallicity of the isochrone by 0.5\,dex decreases the derived $T_{\text{eff}}$ by only $\sim40$\,K.

The standard deviation of the $T_{\text{eff}}$ residuals with respect to \citet{hem+18} is 170\,K.
We adopt this value, subtracted in quadrature by the $T_{\text{eff}}$ measurement uncertainty of 50\,K in \citet{hem+18}, as our systematic uncertainty in $T_{\text{eff}}$.
Propagating this systematic uncertainty to $\log{g}$ results in a $\log{g}$ systematic uncertainty of 0.25\,dex. 
Random uncertainties were derived by propagating photometric uncertainties in the public SkyMapper photometry to these stellar parameter derivations.
The total uncertainty was taken as the quadrature sum of the systematic and random uncertainties.

\subsection{Metallicity Analysis}
\label{sec:metallicity}

\begin{figure}[!htbp]
\centering
\includegraphics[width =\columnwidth]{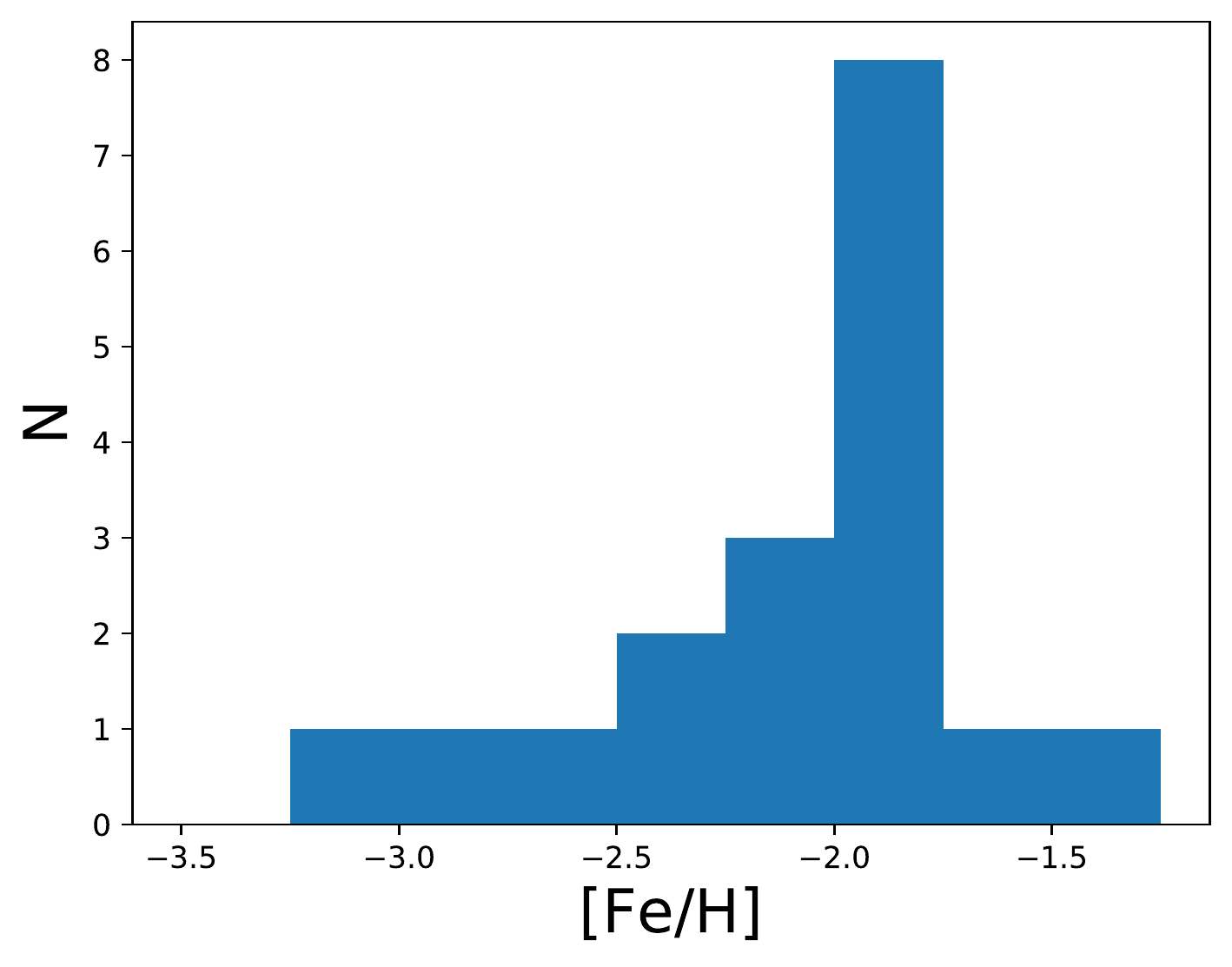}
\caption{Histogram of the metallicities of our sample of 18 our newly discovered members red giant stars.
The distribution peaks just above the very metal-poor regime ([Fe/H] $= -2.0$) with a tail extending to extremely metal-poor ([Fe/H] $= -3.0$) metallicities.}
\label{fig:feh_hist}
\end{figure}

We derived metallicities for the stars in our sample using independently the Ca II K absorption feature ($\sim$3933.7\,\AA), the Mg b absorption region ($\sim$5180\,\AA), and the calcium triplet absorption features ($\sim8500$\,\AA).
The methods for deriving each of these metallicities are described below. 
The metallicities we derive are shown in Table~\ref{tab:measurements}, and sample spectra are shown in Figure~\ref{fig:specs}.
A histogram of our final metallicities is shown in Figure~\ref{fig:feh_hist}.

\subsubsection{Ca II K line}
\label{sec:ca2k}

We follow the calibration of \citet{brn+99} to derive metallicities from the Ca II K absorption line at $\sim3933.7$\,\AA.
This calibration requires $B-V$ colors and a measure of the strength of the Ca II line known as the KP index, which is an estimate of the pseudo-equivalent width of the feature. 
We derive the KP index following exactly the procedures presented in \citet{brn+99}.
The $B-V$ colors of these stars were derived by first transforming SkyMapper $g$ and $r$ photometry to the Pan-STARRS system using a sample photometry of F, G, and K stellar type stars from the SkyMapper website \citep{p+98}.
We then used the transformations in \citet{tsl+12} to convert from Pan-STARRS $g$ and $r$ photometry to $B-V$ colors.

We note that we excluded Ca II K metallicity estimates if the procedure returned a value of [Fe/H] $> -1.5$. This is the metallicity regime where the Ca II K line saturates (see top left panel of Figure~\ref{fig:specs}) and no longer produces accurate metallicity estimates. 
This led to removal of metallicities of three stars: Sgr-62, Sgr-136, and Sgr-141.
Finally, we note that the \citet{brn+99} calibration assumes a [Ca/Fe] = 0.4 when [Fe/H], given that this is a calibration for Milky Way halo stars.
We assume the same for these Sagittarius dSph stars, given the general agreement in the [Ca/Fe] values between halo stars and Sagittarius dSph metal-poor stars found by \citet{hem+18}.
Uncertainties in the metallicities of these stars were derived as the quadrature sum of the systematic uncertainty provided in \citet{brn+99}, and the random uncertainties propagated by shifting the continuum and from propagating the uncertainty in the $B-V$ color.

\subsubsection{Mg b line triplet lines}
\label{sec:mg}

Metallicities were derived from the Mg b line region (5150\,{\AA} to 5190\,{\AA}) via standard spectral synthesis techniques, following \citet{cf+19}.
We do note that several previous studies have used the Mg b region to derive metallicities of stars in dwarf galaxies \citep{sdl+15,wom+15, wmo+16}.
Specifically, synthetic spectra were generated using the 2017 version of the MOOG 1D LTE radiative transfer code \citep{s+73, sks+11}, the Kurucz model atmospheres \citep{ck+04}, and a linelist combining data from \citet{k+11} and \citet{slc+09, slr+14, sck+16}.
The [Mg/Fe] in these syntheses was set as [Mg/Fe] = 0.3, matching the general [$\alpha$/Fe] trend for metal-poor stars in Sagittarius  \citep{hem+18}.
These syntheses were performed within the SMH software \citep{c+14}, which enabled the visual identification of the synthetic spectrum that best matched each observed spectrum.
The [Fe/H] of the best matching synthetic spectrum was taken as the [Fe/H] of the observed spectrum.

The random uncertainty from the fitting procedure was assumed to be the difference in [Fe/H] when requiring to encompass the noise in the observed spectrum with synthetic spectra. 
The systematic uncertainty was determined by re-deriving metallicities after shifting the T$_{\rm eff}$ and $\log g$ values by their $1\sigma$ uncertainties.
The total uncertainty was assumed as the quadrature sum of the random and systematic uncertainties.
We note that additional uncertainty from variations in the [Mg/Fe] is likely far less than our total uncertainties ($\sim0.35$\,dex).
Sagittarius stars with [Fe/H] $< -1.5$ in \citet{hem+18} have a scatter of $\sim0.15$\,dex in their [Ca/Fe].
Since Mg, like Ca, is an $\alpha$-element, this suggests a similarly low scatter in the Mg abundances relative to our uncertainties.

\subsubsection{Calcium triplet lines}
\label{sec:CaT}

We derived metallicities from the calcium triplet lines using the calibration of \citet{cpg+13}.
This calibration relates the metallicity of a star to its absolute $V$ magnitude and the sum of the equivalent widths of the three calcium triplet lines at 8498\,{\AA}, 8542\,{\AA}, and 8662\,{\AA}.
We calculated these equivalent widths by fitting the Voigt1D model in the python astropy package \citep{astropy} to each line.
Absolute $V$ magnitudes were derived following the color transformations described in Section~\ref{sec:ca2k}, along with an updated distance modulus of the Sagittarius dSph (17.10, \citealt{fs+20}).
Random uncertainties were determined by re-deriving the metallicity after varying the continuum by $1\sigma$, as determined by the signal-to-noise of the spectrum.
Systematic uncertainties were assumed to be 0.17\,dex following \citet{cpg+13}.
Total uncertainties were derived as the quadrature sum of the random and systematic uncertainties.

\subsubsection{Final metallicity values \& validation}
\label{sec:finalfeh}

We derived final metallicity values by taking the average of the three metallicity estimators (see Sections~\ref{sec:ca2k} to~\ref{sec:CaT}), weighted by the inverse square of their uncertainties.
The final uncertainty was taken as the uncertainty in the weighted average.
Final metallicities and uncertainties are presented in Table~\ref{tab:measurements}.

To validate our metallicities, we also derived metallicities for four metal-poor giant stars (HD21581, HD216143, HD122563, and CS22892-052) that we observed with the same MagE observational setup.
We obtained metallicities for these stars following the aforementioned methods, and obtained the following values: 
[Fe/H] = $-1.60\pm0.15$ for HD21581, 
[Fe/H] = $-2.25\pm0.15$ for HD216143, 
[Fe/H] = $-2.67\pm0.13$ for HD122563, and 
[Fe/H] = $-3.01\pm0.14$ for CS22892-052.
These metallicities all agree within 1$\sigma$ of the literature metallicities of these stars, which are 
[Fe/H] = $-1.70$ for HD21581 \citep{rpt+14}, 
[Fe/H] = $-2.15$ for HD216143 \citep{bg+16}, 
[Fe/H] = $-2.79$ for HD122563 \citep{fcj+13}, and 
[Fe/H] = $-3.08$ for CS22892-052 \citep{fcj+13}.

\begin{figure*}[!htbp]
\centering
\includegraphics[width=1.00\textwidth]{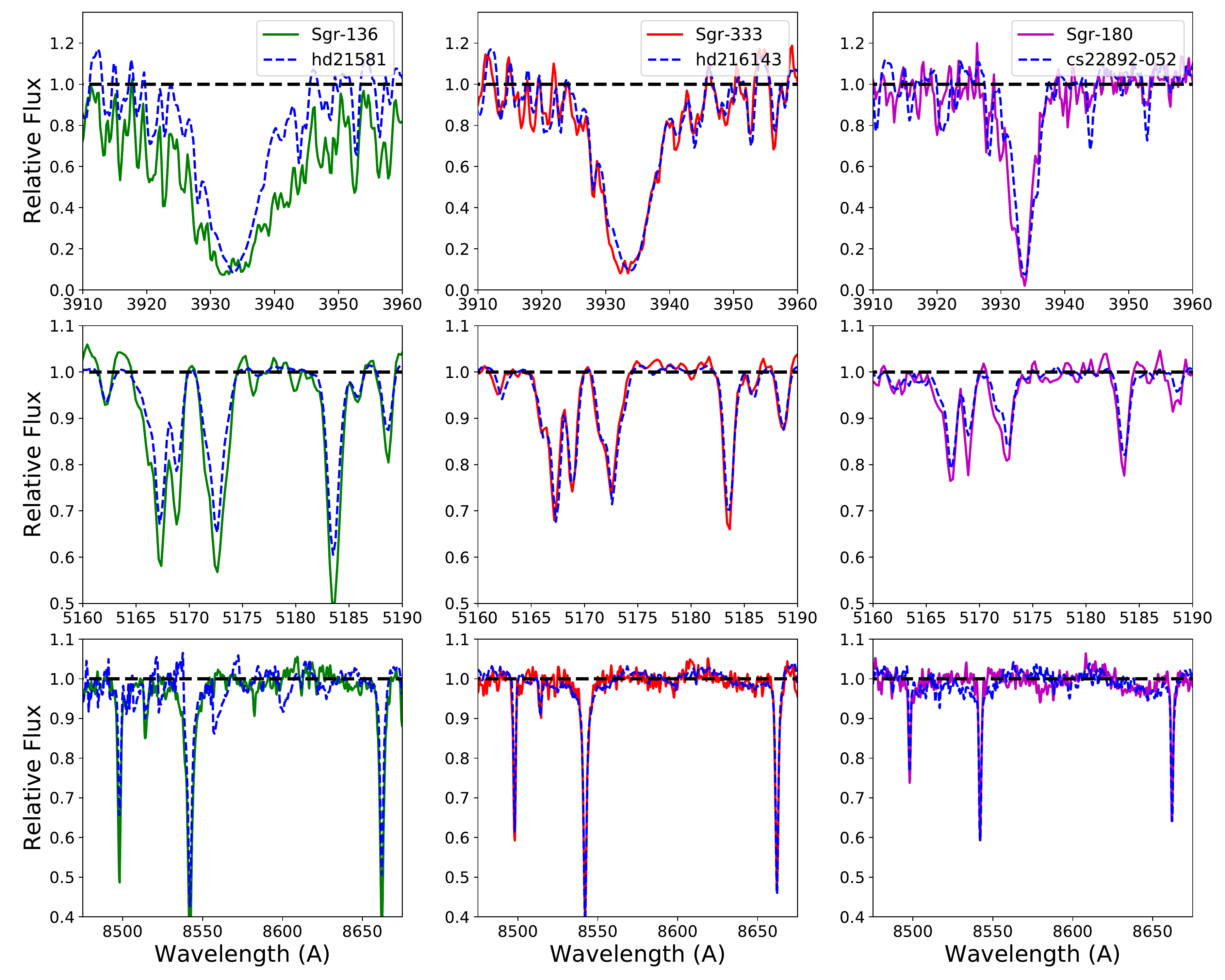}
\caption{Sample spectra over the Ca II K line at 3933.7\,{\AA} (top panels), the Mg b line region at $\sim5150$\,{\AA} (middle panels), and the calcium triplet lines around $\sim$8550\,{\AA} (bottom panels). MagE spectra of Sgr-136 ([Fe/H] = $-1.69$) and HD21581 ([Fe/H] = $-1.56$; \citealt{rpt+14}) are shown on the left panels, Sgr-333 ([Fe/H] = $-2.10$) and HD216143 ([Fe/H] = $-2.24$; \citealt{bg+16}) on the middle panels, and Sgr-180 ([Fe/H] = $-3.08$) and CS22892-052 ([Fe/H] = $-3.08$; \citealt{fcj+13}) on the right panels.
HD21581 and CS22892-052 have slightly weaker absorption features than Sgr-136 and Sgr-180 due to their higher ($\sim300$\,K) effective temperatures.
HD216143 and Sgr-333 have absorption features of similar strengths, due to their proximate metallicities and effective temperatures.}
\label{fig:specs}
\end{figure*}

In Figure~\ref{fig:specs}, we  provide a visual comparison of the spectra of three of these metal-poor giants to selected Sagittarius members with similar stellar parameters and metallicities.
As can be seen, the absorption lines of HD21581 and CS22892-052 are slightly weaker than Sgr-136 and Sgr-180, respectively, despite their similar metallicities. This is due to HD21581 and CS22892-052 being on average $\sim300$\,K warmer than the two Sagittarius members.
HD216143 and Sgr-333 have nearly identical absorption features and effective temperatures (4600\,K vs. 4546\,K), validating our derivation of similar metallicities ([Fe/H] = $-2.24$ vs. [Fe/H] = $-2.10$) for both stars.

%%%%%%%%%%%%%%%%%%%%%%%%%%%%%%%%%%%%%%%%%%%%%%%%%%%%%%%%%%%%%%%%%%%%%%%%%%%%%%%%%
%%%%%%%%%%%%%%%%%%%%%%%%%%%%%%%%%%%%%%%%%%%%%%%%%%%%%%%%%%%%%%%%%%%%%%%%%%%%%%%%%

\subsection{Carbon Abundances} \label{sec:C}

\begin{figure*}[!htbp]
\centering
\includegraphics[width =1.00\textwidth]{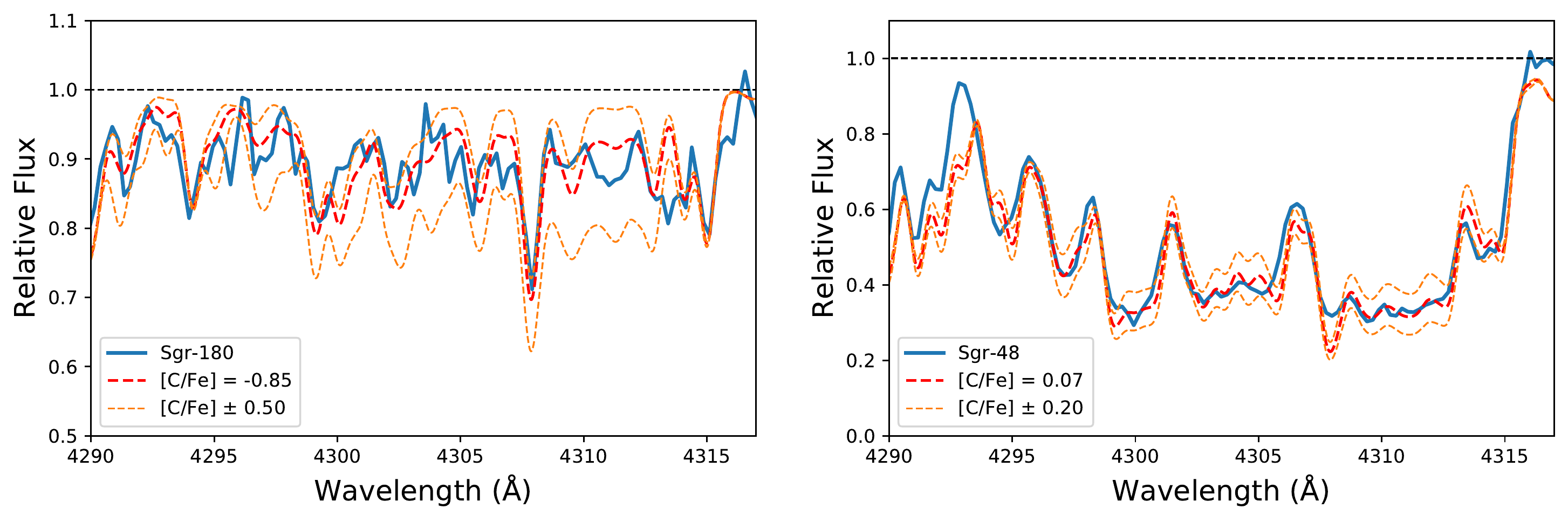}
\caption{Left: The CH G bandhead region of our most metal-poor star, Sgr-180 (blue), compared to its best-matching synthetic spectra (red). 
Synthetic spectra with carbon abundances offset by $\pm 0.50$ are shown as dashed orange lines, and the continuum is marked as a dashed black line to guide the eye.
Right: The same plot but with our most carbon-enhanced star, Sgr-48, shown in blue. 
The best matching synthetic spectrum is again shown in red, and synthetic spectra with carbon abundances offset by $\pm 0.20$ are shown as dashed orange lines. }
\label{fig:carbon}
\end{figure*}

We also derived a carbon abundance ([C/Fe]) for each of our red giants from the CH $G$ bandhead region (4275\,{\AA} to 4320\,{\AA}) using standard spectral synthesis techniques, analogous to our derivation of metallicities from the Mg b line region in Section~\ref{sec:mg}.
The line list for these syntheses included CH molecular line data from \citet{mpv+14} in addition to the sources listed in Section~\ref{sec:mg}.
The [C/Fe] was varied to find the best fitting synthetic spectrum while the [Fe/H] was set to each star's final metallicity value.
The random uncertainty in [C/Fe] was taken as the variation in [C/Fe] needed to encompass the noise in the G band region.
The systematic uncertainties were determined by re-deriving [C/Fe] after varying T$_{\rm eff}$, $\log g$, and the metallicity of each star by their 1$\sigma$ uncertainties.
All these sources of uncertainty were added in quadrature to derive a total uncertainty.
Examples of our carbon syntheses are shown in Figure~\ref{fig:carbon}.
In Table~\ref{tab:measurements}, we present our derived carbon abundances in addition to corrected carbon abundances that account for the depletion of the surface carbon abundance as stars ascend the red giant branch \citep{pfb+14}.

%%%%%%%%%%%%%%%%%%%%%%%%%%%%%%%%%%%%%%%%%%%%%%%%%%%%%%%%%%%%%%%%%%%%%%%%%%%%%%%%%
%%%%%%%%%%%%%%%%%%%%%%%%%%%%%%%%%%%%%%%%%%%%%%%%%%%%%%%%%%%%%%%%%%%%%%%%%%%%%%%%%

%%%%%%%%%%%%%%%%%%%%%%%%%%%%%%%%%%%%%%%%%%%%%%%%%%%%%%%%%%%%%%%%%%%%%%%%%%%%%%%%%
\begin{deluxetable*}{rrrrrrrrr}
\tablecaption{\label{tab:measurements} Stellar parameters and chemical abundances}
\tablehead{
\colhead{Name} & 
  \colhead{$T_{\text{eff}}$} & 
  \colhead{log\,$g$} &
\colhead{[Fe/H]$_{\text{Mg}}$} & 
\colhead{[Fe/H]$_{\text{CaT}}$} & 
\colhead{[Fe/H]$_{\text{CaIIK}}$} & 
\colhead{[Fe/H]$_{\text{final}}$} & 
\colhead{[C/Fe]} & 
\colhead{[C/Fe]$_{\text{corrected}}$ $\tablenotemark{a}$}\\
  \colhead{} &
  \colhead{(K)} & 
  \colhead{(dex)} &
  \colhead{(dex)} & 
  \colhead{(dex)} & 
  \colhead{(dex)} & 
  \colhead{(dex)} & 
  \colhead{(dex)} & 
  \colhead{(dex)}}
\startdata
Sgr-62 & $4530\pm170$ & $1.11\pm0.30$ & $-1.40\pm0.36$ & $-1.49\pm0.22$ & \nodata & $-1.47\pm0.19$ & $-0.53\pm0.31$ & $0.01\pm0.31$ \\
Sgr-136 & $4400\pm160$ & $0.88\pm0.25$ & $-1.61\pm0.37$ & $-1.71\pm0.17$ & \nodata & $-1.69\pm0.15$ & $-0.73\pm0.30$ & $-0.08\pm0.30$ \\
Sgr-141 & $4630\pm160$ & $1.30\pm0.25$ & $-1.89\pm0.33$ & $-1.74\pm0.18$ & \nodata & $-1.77\pm0.16$ & $-0.51\pm0.33$ & $0.04\pm0.33$ \\
Sgr-157 & $5110\pm180$ & $2.26\pm0.30$ & $-2.00\pm0.39$ & $-1.80\pm0.20$ & $-1.64\pm0.43$ & $-1.81\pm0.16$ & $-0.37\pm0.34$ & $-0.36\pm0.34$ \\
Sgr-162 & $4660\pm200$ & $1.37\pm0.35$ & $-1.96\pm0.41$ & $-1.79\pm0.24$ & $-1.66\pm0.59$ & $-1.81\pm0.20$ & $-0.36\pm0.39$ & $0.13\pm0.39$ \\
Sgr-265 & $4450\pm160$ & $0.97\pm0.25$ & $-1.88\pm0.37$ & $-1.73\pm0.20$ & $-2.05\pm0.32$ & $-1.83\pm0.15$ & $-0.63\pm0.31$ & $0.02\pm0.31$ \\
Sgr-225 & $4520\pm160$ & $1.09\pm0.25$ & $-2.02\pm0.35$ & $-1.85\pm0.18$ & $-1.52\pm0.43$ & $-1.84\pm0.15$ & $-0.72\pm0.29$ & $-0.08\pm0.29$ \\
Sgr-48 & $4310\pm160$ & $0.69\pm0.25$ & $-2.02\pm0.37$ & $-1.81\pm0.18$ & $-2.09\pm0.25$ & $-1.92\pm0.14$ & $0.07\pm0.21$ & $0.56\pm0.21$ \\
Sgr-182 & $5140\pm170$ & $2.33\pm0.25$ & $-2.14\pm0.35$ & $-1.95\pm0.23$ & $-1.76\pm0.37$ & $-1.95\pm0.17$ & $-0.44\pm0.38$ & $-0.43\pm0.38$ \\
Sgr-91 & $4680\pm160$ & $1.40\pm0.25$ & $-2.12\pm0.35$ & $-1.98\pm0.19$ & $-1.94\pm0.43$ & $-2.00\pm0.16$ & $-0.41\pm0.31$ & $0.12\pm0.31$ \\
Sgr-298 & $5170\pm190$ & $2.40\pm0.30$ & $-2.15\pm0.33$ & $-2.03\pm0.20$ & $-1.96\pm0.36$ & $-2.04\pm0.15$ & $-0.18\pm0.40$ & $-0.17\pm0.40$ \\
Sgr-333 & $4550\pm160$ & $1.16\pm0.25$ & $-2.30\pm0.31$ & $-2.08\pm0.18$ & $-1.85\pm0.44$ & $-2.10\pm0.15$ & $-0.93\pm0.34$ & $-0.25\pm0.34$ \\
Sgr-300 & $4440\pm160$ & $0.95\pm0.25$ & $-2.17\pm0.35$ & $-2.12\pm0.18$ & $-2.33\pm0.20$ & $-2.21\pm0.12$ & $-0.65\pm0.30$ & $0.11\pm0.30$ \\
Sgr-81 & $4870\pm170$ & $1.77\pm0.25$ & $-2.53\pm0.34$ & $-2.18\pm0.23$ & $-2.57\pm0.29$ & $-2.37\pm0.16$ & $0.26\pm0.29$ & $0.48\pm0.29$ \\
Sgr-198 & $4450\pm160$ & $0.97\pm0.25$ & $-2.45\pm0.32$ & $-2.40\pm0.21$ & $-2.59\pm0.26$ & $-2.47\pm0.15$ & $-0.98\pm0.33$ & $-0.20\pm0.33$ \\
Sgr-69 & $4380\pm160$ & $0.83\pm0.25$ & $-2.79\pm0.39$ & $-2.43\pm0.20$ & $-2.64\pm0.17$ & $-2.58\pm0.12$ & $-0.73\pm0.33$ & $0.04\pm0.33$ \\
Sgr-38 & $4680\pm160$ & $1.40\pm0.25$ & $-2.77\pm0.36$ & $-2.76\pm0.20$ & $-2.95\pm0.27$ & $-2.82\pm0.15$ & $-0.30\pm0.43$ & $0.23\pm0.43$ \\
Sgr-180 & $4540\pm160$ & $1.14\pm0.25$ & $-2.98\pm0.37$ & $-3.07\pm0.21$ & $-3.13\pm0.22$ & $-3.08\pm0.14$ & $-0.85\pm0.47$ & $-0.11\pm0.47$ \\
\hline
HD21581 & 4940$\tablenotemark{b}$ & 2.10$\tablenotemark{b}$ & $-1.85\pm0.32$ & $-1.55\pm0.18$ & \nodata & $-1.60\pm0.15$ & \nodata & \nodata \\
HD216143 & 4530$\tablenotemark{c}$ & 1.10$\tablenotemark{c}$ & $-2.42\pm0.32$ & $-2.22\pm0.19$  & $-2.14\pm0.36$ & $-2.25\pm0.15$ & \nodata & \nodata \\
HD122563 & 4612$\tablenotemark{d}$ & 0.85$\tablenotemark{d}$ & $-2.70\pm0.34$ & $-2.60\pm0.18$  & $-2.78\pm0.25$ & $-2.67\pm0.13$ & \nodata & \nodata \\
CS22892-052 & 4828$\tablenotemark{d}$ & 1.35$\tablenotemark{d}$ & $-3.11\pm0.37$ & $-2.91\pm0.19$  & $-3.12\pm0.23$ & $-3.01\pm0.14$ & \nodata & \nodata \\
\enddata
\tablenotetext{a}{Corrected for the evolutionary state of the star following \citet{pfb+14}.}
\tablenotetext{b}{Taken from \citet{rpt+14}.}
\tablenotetext{c}{Taken from \citet{bg+16}.}
\tablenotetext{d}{Taken from \citet{fcj+13}.}
\end{deluxetable*}
%%%%%%%%%%%%%%%%%%%%%%%%%%%%%%%%%%%%%%%%%%%%%%%%%%%%%%%%%%%%%%%%%%%%%%%%%%%%%%%%%

%%%%%%%%%%%%%%%%%%%%%%%%%%%%%%%%%%%%%%%%%%%%%%%%%%%%%%%%%%%%%%%%%%%%%%%%%%%%%%%%%
%%%%%%%%%%%%%%%%%%%%%%%%%%%%%%%%%%%%%%%%%%%%%%%%%%%%%%%%%%%%%%%%%%%%%%%%%%%%%%%%%

\section{Results and Conclusions} 
\label{sec:results}

We present the metallicities and carbon abundances of eighteen red giant stars in the Sagittarius dSph that were identified as metal-poor candidates with publicly available, metallicity-sensitive photometry from SkyMapper DR1.1 \citep{wol+18}.
Notably, eight of these stars are very metal-poor ($-3.0 < $ [Fe/H] $\leq -2.0$) and one is extremely metal-poor ([Fe/H] $\leq -3.0$), more than doubling the known $\sim5$ very metal-poor stars in the system (e.g., \citealt{hem+18, cf+19}) and identifying the one of the first known extremely metal-poor stars in the Sagittarius dSph\textsuperscript{\ref{ft:sgr}}.
This result conclusively shows that even the most massive satellite dwarf spheroidal galaxy, Sagittarius, has a metallicity distribution function extending to the extremely metal-poor regime as is also seen in other, less massive dwarf spheroidal galaxies (e.g., \citealt{fks+10, tjl+19}).

Our detection of these very metal-poor stars in the Sagittarius dSph aligns with theoretical expectations that all galaxies should plausibly host chemically primitive stellar populations. 
These stars would likely be remnants from early generations of star formation (e.g., \citealt{dbk+15}), or could plausibly originate from smaller, more chemically primitive dwarf galaxies that were accreted onto the system, both of which are processes that should occur in the formation of larger dwarf spheroidal systems.
The previous scarcity of known stars with [Fe/H] $< -2.0$ in the Sagittarius dSph was then likely caused by its dominant stellar population having a peak metallicity of [Fe/H] $\sim -0.5$ \citep{mbi+17}, which would render very metal-poor stars in the system relatively rare.
With our newly discovered sample of very metal-poor stars, we can now investigate the early chemical evolution of this system and compare it with other galaxies.

\begin{figure}[tbp]
\centering
\includegraphics[width =\columnwidth]{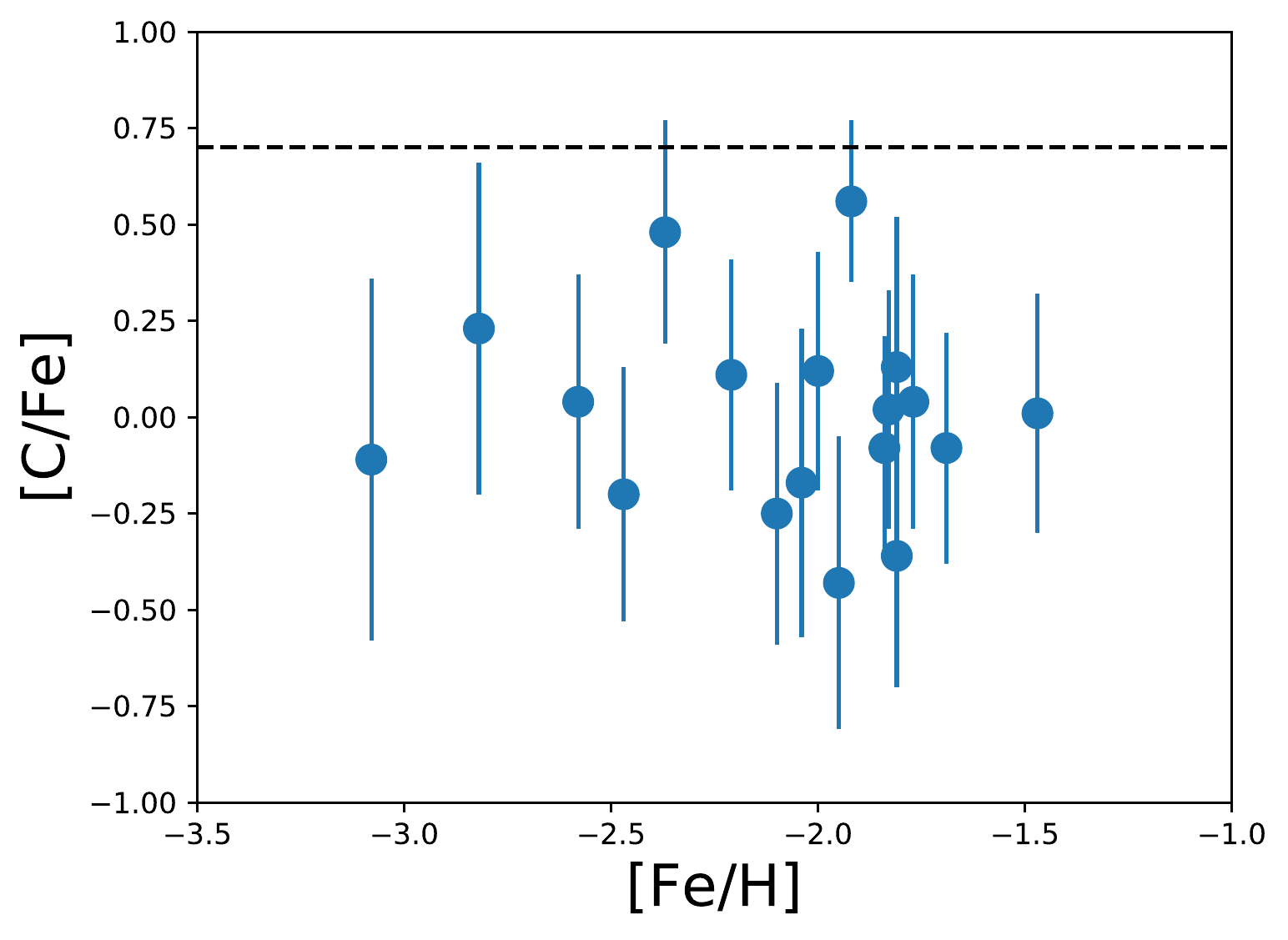}
\caption{Carbon abundances as a function of [Fe/H] for stars in our sample. The plotted carbon abundances have been corrected for the evolutionary state of the star following \citet{pfb+14}. 
The dashed line indicates a carbon enhancement of [C/Fe] = 0.7, above which value stars are defined as carbon-enhanced metal-poor (CEMP) stars.
}
\label{fig:cmet}
\end{figure}

One curious observed signature of our sample is that none of the stars can be classified as carbon-enhanced metal-poor (CEMP; [C/Fe] $> 0.7$) stars (see Figure~\ref{fig:cmet}).
In contrast, one prominent signature among old metal-poor stars in the Milky Way's halo is the  increase of relative carbon enhancement with decreasing metallicity.
Around 20\% of stars in the halo are classified as CEMP stars when [Fe/H] $< -2.0$, and 43\% of stars are CEMP when [Fe/H] $< -3.0$ \citep{pfb+14}.
Combining our sample with the sample in \citet{hem+18} and \citet{cf+19} results in 14 stars with [Fe/H] $\leq -2.0$ in the Sagittarius dSph, none of which are CEMP stars.
There is a 4\% probability of observing no CEMP stars in a sample of 14 with [Fe/H] $\leq -2.0$, if the Sagittarius dSph had the same CEMP fraction as the halo.
This probability hints that the Sagittarius dSph may have a lower CEMP fraction than the Milky Way halo in the very metal-poor regime.

%%%%%%%%%%%%%%%%%%%%%%%%%%%%%%%%%%%%%%%%%%%%%%%%%%%%%%%%%%%%%%%%%%%%%%%%%%%%%%%%%
%%%%%%%%%%%%%%%%%%%%%%%%%%%%%%%%%%%%%%%%%%%%%%%%%%%%%%%%%%%%%%%%%%%%%%%%%%%%%%%%%

This possible discrepancy between the CEMP fractions in the Sagittarius dSph and the Milky Way halo may hint at some dependence of early chemical evolution on the environment in which stars form.
The CEMP fraction in other dwarf spheroidal galaxies also appears to be lower than the halo CEMP fraction when [Fe/H] $\leq -2.0$ (e.g., Carina, Draco, Sculptor; \citealt{vsi+12, kgz+15}), although it may again increase when [Fe/H] $\leq -3.0$ (e.g., \citealt{csf+18}). 
Further targeted studies of extremely metal-poor stars would be helpful in further investigating this trend.
A similar lack of CEMP stars is observed in the Galactic bulge (e.g., \citealt{hca+15, hak+16}), also suggesting that early chemical evolution may not be universal. 
However, these discrepancies do not invalidate the Milky Way halo being assembled from the accretion of smaller galaxies, as the spread in carbon abundances of stars in the Milky Way halo may originate from a variety of galaxies that assembled to form the halo. 

We note for completion that metallicities derived from the SkyMapper $v$ filter are biased high for carbon-enhanced stars, which can lead to them being excluded from our sample and thus artificially decrease any CEMP fraction \citep{dbm+19, cfj+20}.
However, we emphasize that this selection effect should be negligible for our sample discussed here. Due to the weakening precision of the $v$ band photometry in SkyMapper DR1.1 at the magnitudes of these stars ($g \sim 15$ to $g \sim 17.5$), our photometric metallicities had large ($\sim 0.75$\,dex) uncertainties. 
Any bias in the photometric metallicities of stars at the CEMP threshold of [C/Fe] = 0.7, after carbon-correction following \citealt{pfb+14}, would have been lower than these uncertainties \citep{cfj+20}.
Accordingly, the large uncertainties in the photometric metallicities in our selection procedure would supercede much of the bias against CEMP stars.
We also note that the metallicity distribution of our observed stars peaks above [Fe/H] $= -2.0$, suggesting that our selection function does select stars at higher metallicities than [Fe/H] = $-2.0$.
Accordingly, this suggests that stars with [Fe/H] $< -2.0$ with slightly artificially higher photometric metallicities due to carbon enhancement would still have been selected for our sample. 
As a result, our lack of detected CEMP stars is likely independent of our target selection procedure.

\begin{figure*}[!htbp]
\centering
\includegraphics[width =1.00\textwidth]{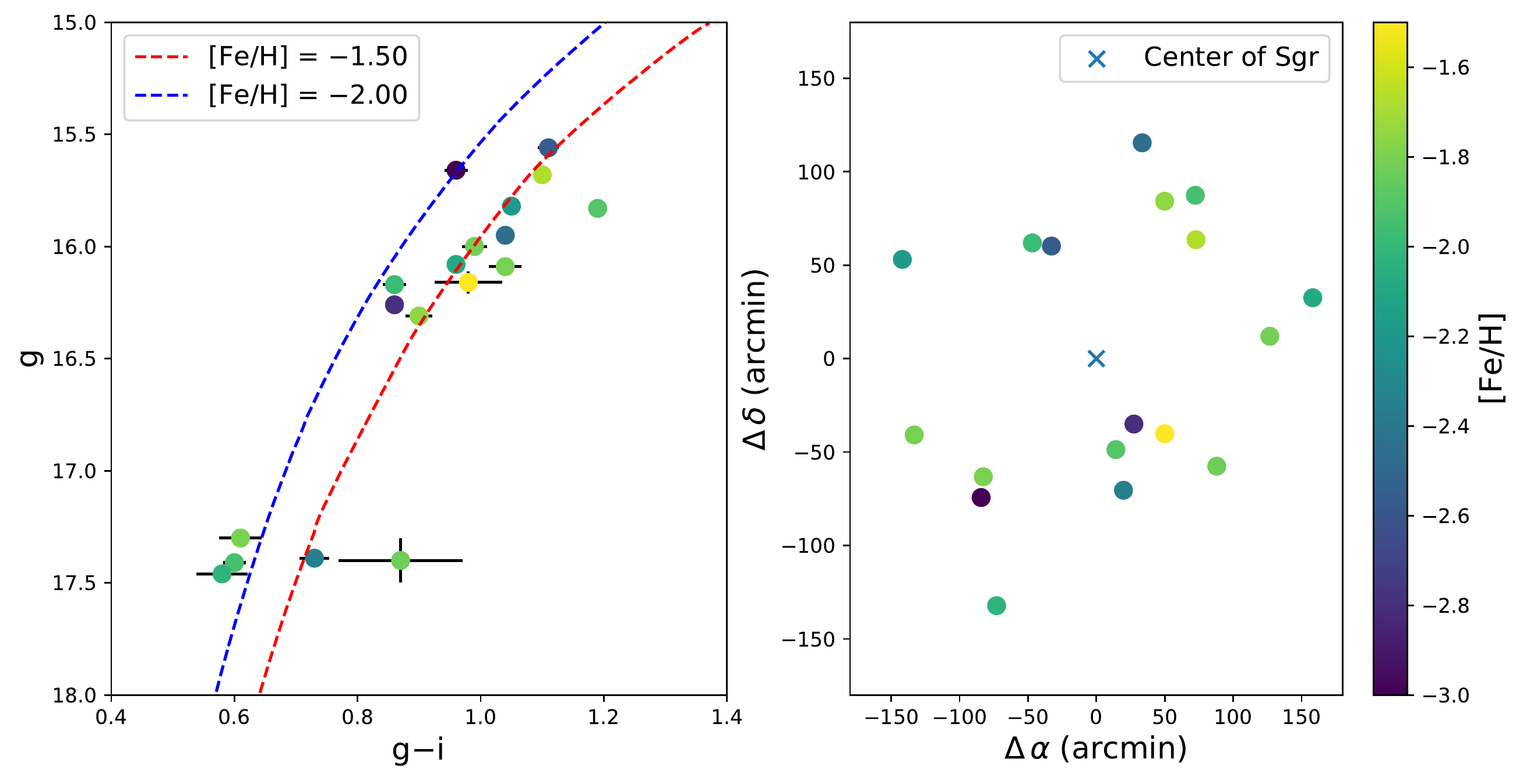}
\caption{Left: Color-magnitude diagram of our observed Sagittarius stars colored by their metallicities. Two 12\,Gyr isochrones with [Fe/H] = $-2.0$ and [Fe/H] $= -1.5$ from the MESA Isochrones \& Stellar Tracks database \citep{d+16, cdc+16, pbd+11, pca+13, pms+15, psb+18} database are overlaid at the distance modulus of the Sagittarius dSph (17.10, \citealt{fs+20}).
Right: Position of our Sagittarius members with respect to the center of the Sagittarius dSph, which is marked with a blue cross.}
\label{fig:spatial}
\end{figure*}

At a broader level, we demonstrate that dedicated, wide-field searches for the most metal-poor stars in large dwarf galaxies are feasible using public metallicity-sensitive photometry.
As shown in Figure~\ref{fig:spatial}, all of our observed Sagittarius dSph stars are notably distant from the nucleus of the system ($r_h = 0.43'\pm0.08'$; \citealt{bic+08}), but lie within its main body ($r_c = 224'\pm12'$; \citealt{msw+03}).
Searches for the most metal-poor stars in the outskirts of dwarf spheroidal galaxies could be particularly productive, since at least some of these systems are known to have metallicity gradients (e.g., \citealt{eih+04}).
At face value, we unfortunately cannot interpret the spatial distribution of our Sagittarius dSph stars further to investigate, for instance, a metallicity gradient, given our relatively small sample of stars and qualitative selection function.
However, such work and more precise targeting of the most metal-poor stars will be possible with the improved photometric precision in future SkyMapper data releases (e.g., \citealt{owb+19}).
At minimum, future high-resolution spectroscopy of these stars will derive their detailed chemical abundance patterns. 
Such work will enable comprehensive studies of the early chemical evolution and formation history of the massive systems that were accreted to form Milky Way halo.

%%%%%%%%%%%%%%%%%%%%%%%%%%%%%%%%%%%%%%%%%%%%%%%%%%%%%%%%%%%%%%%%%%%%%%%%%%%%%%%%%
%%%%%%%%%%%%%%%%%%%%%%%%%%%%%%%%%%%%%%%%%%%%%%%%%%%%%%%%%%%%%%%%%%%%%%%%%%%%%%%%%

\acknowledgements

A.C. and A.F. acknowledge support from NSF grant AST-1716251.  
We thank Alexander Ji for providing a literature compilation of dwarf galaxy stars.
This work made use of NASA's Astrophysics Data System Bibliographic Services, and the SIMBAD database, operated at CDS, Strasbourg, France \citep{woe+00}.

This work has made use of data from the European Space Agency (ESA) mission
{\it Gaia} (\url{https://www.cosmos.esa.int/gaia}), processed by the {\it Gaia}
Data Processing and Analysis Consortium (DPAC,
\url{https://www.cosmos.esa.int/web/gaia/dpac/consortium}). Funding for the DPAC
has been provided by national institutions, in particular the institutions
participating in the {\it Gaia} Multilateral Agreement.

The national facility capability for SkyMapper has been funded through ARC LIEF grant LE130100104 from the Australian Research Council, awarded to the University of Sydney, the Australian National University, Swinburne University of Technology, the University of Queensland, the University of Western Australia, the University of Melbourne, Curtin University of Technology, Monash University and the Australian Astronomical Observatory. SkyMapper is owned and operated by The Australian National University's Research School of Astronomy and Astrophysics. The survey data were processed and provided by the SkyMapper Team at ANU. The SkyMapper node of the All-Sky Virtual Observatory (ASVO) is hosted at the National Computational Infrastructure (NCI). Development and support the SkyMapper node of the ASVO has been funded in part by Astronomy Australia Limited (AAL) and the Australian Government through the Commonwealth's Education Investment Fund (EIF) and National Collaborative Research Infrastructure Strategy (NCRIS), particularly the National eResearch Collaboration Tools and Resources (NeCTAR) and the Australian National Data Service Projects (ANDS).

Facilities: Magellan-Baade \citep[MagE;][]{mbt+08}, SkyMapper \citep{ksb+07}

\software{MOOG \citep{s+73, sks+11}, MagE CarPy \citep{k+03}, Astropy \citep{astropy}, NumPy \citep{numpy}, SciPy \citet{jop+01}, Matplotlib \citep{Hunter+07}}

\appendix

\section{Velocities of objects omitted from metallicity analysis}

In Table~\ref{tab:nonmem}, we present the coordinates, magnitudes, and velocities of objects that were observed in our program but were omitted from Table~\ref{tab:obs}.
Sixteen of these were omitted as they were classified as non-members of Sagittarius, as determined from their radial velocities (see Section~\ref{sec:rvs}). 
Three more of these objects had velocities consistent with membership, but were excluded from further analysis to their being e.g., a spectroscopic binary, or having distorted hydrogen Balmer lines. 
These details are provided in the comments column of Table~\ref{tab:nonmem}.
Similar to the sample of stars presented in Table~\ref{tab:obs}, the uncertainty on the velocities of these stars is $\sim$3\,km/s.

%%%%%%%%%%%%%%%%%%%%%%%%%%%%%%%%%%%%%%%%%%%%%%%%%%%%%%%%%%%%%%%%%%%%%%%%%%%%%%%%%
\begin{deluxetable*}{lllrrrrrl}[!htbp] % <--- column justification (center/left/right)
\tablecaption{\label{tab:nonmem} Velocities of objects omitted from metallicity analysis}
\tablehead{   % column headings
  \colhead{Name} &
  \colhead{RA (h:m:s)} & 
  \colhead{DEC (d:m:s)} &
%  \colhead{Slit size} &
  \colhead{$g$ } &
  \colhead{$v_{\text{helio}}$ } &
  \colhead{Comments}\\
   \colhead{}&
   \colhead{(J2000)}&
   \colhead{(J2000)}&
%   \colhead{}&
   \colhead{[mag]}&
   \colhead{[km\,s$^{-1}$]}&
   \colhead{}
  }
\startdata
Sgr-317 & 18:43:33.8001 & $-$31:14:41.5383 & 16.02 & $-$1 & \\
Sgr-105 & 18:49:00.7118 & $-$30:55:59.5389 & 15.52 & 181 & \\
Sgr-87 & 18:49:38.4731 & $-$30:17:13.1231 & 14.96 & 135 & Has Balmer emission lines\\
Sgr-68 & 18:50:11.9385 & $-$30:50:06.6329 & 16.08 & 10 & \\
Sgr-134 & 18:50:35.179 & $-$31:47:22.3161 & 15.89 & 33 & \\
Sgr-179 & 18:51:11.3552 & $-$28:54:55.0119 & 16.60 & $-$132 & \\
Sgr-186 & 18:51:57.8331 & $-$32:18:49.8092 & 15.79 & 74 & \\
Sgr-349 & 18:52:21.1407 & $-$27:51:24.9048 & 16.46 & $-$58 & \\
Sgr-127 & 18:55:53.378 & $-$32:02:01.2355 & 15.93 & 25 & \\
Sgr-205 & 18:56:20.4976 & $-$28:31:17.5376 & 16.18 & $-$92 & \\
Sgr-177 & 18:58:29.6478 & $-$28:50:05.0311 & 15.86 & 29 & \\
Sgr-249 & 19:01:22.094 & $-$32:24:52.5733 & 17.67 & 177 & \\
Sgr-148 & 19:01:29.765 & $-$31:36:43.6175 & 17.26 & 45 & \\
Sgr-7 & 19:01:43.3463 & $-$29:08:25.4097 & 14.98 & 32 & \\
Sgr-115 & 19:01:48.7458 & $-$30:54:30.8118 & 15.96 & 101 & \\
Sgr-170 & 19:03:29.5049 & $-$31:01:38.7417 & 16.38 & 10 & \\
Sgr-228 & 19:05:16.2722 & $-$30:11:05.3032 & 15.97 & 132 & Distorted Balmer lines\\
Sgr-312 & 19:05:55.7122 & $-$31:48:48.3383 & 18.04 & 126 & Spectroscopic binary\\
Sgr-335 & 19:07:20.3496 & $-$29:46:41.0352 & 16.15 & 188 & \\
\enddata
\end{deluxetable*}
%%%%%%%%%%%%%%%%%%%%%%%%%%%%%%%%%%%%%%%%%%%%%%%%%%%%%%%%%%%%%%%%%%%%%%%%%%%%%%%%%

\bibliography{sgr}

\end{document}